%% file: smf.tex
\documentclass[usenatbib]{mn2e}
\usepackage{multirow}

\newcommand{\apjl}{ApJL}
\newcommand{\aap}{A\&A}

\newcommand{\apjs}{ApJS}

\usepackage{epsfig,amsmath,enumerate,amssymb,aas_macros,bm,booktabs,multirow,xcolor}
\usepackage{graphics}
\input{macros_xie}

\usepackage{multirow}
\usepackage{natbib}
\usepackage{times}

\title{Assembly history of subhalo populations in galactic and cluster  sized dark haloes}
\author [Lizhi Xie, L. Gao]
 {Lizhi Xie$^{1,3}$ \thanks{Email:lzxie@oats.inaf.it}, 
    Liang Gao$^{1,2}$
\\
$^1$The Partner Group of Max Planck Institute for Astrophysics,
National Astronomical Observatories, Chinese Academy of Sciences, Beijing, 100012, China\\
$^2$Institute of Computational Cosmology, Department of Physics,
University of Durham, Science Laboratories, South Road, Durham DH1
3LE \\
$^3$ INAF–Astronomical Observatory of Trieste, via Riccardo Bozzoni,2 , 34124 Trieste, Italy
}

\begin{document}
\label{first page} \maketitle

\begin{abstract}
We make use of two suits of ultra high resolution N-body simulations
of individual dark matter haloes from the Phoenix and the Aquarius
Projects to investigate systematics of assembly history of subhaloes
in dark matter haloes differing by a factor of $1000$ in the halo
mass. We have found that {\it real} progenitors
which built up present day subhalo population are relatively more
abundant for high mass haloes, in contrast to previous studies
claiming a universal form independent of the host halo mass. That is
mainly because of repeated counting of the 're-accreted' (progenitors
passed through and were later re-accreted to the host more than once)
and inclusion of the 'ejected' progenitor population(progenitors were
accreted to the host in the past but no longer members at present day)
in previous studies. The typical accretion time for all progenitors
vary strongly with the host halo mass, which is typical about 
$z \sim 5$ for the galactic Aquarius and about $z \sim 3$ for the 
cluster sized Phoenix haloes. Once these progenitors start to orbit 
their parent haloes, they rapidly lose their original mass but not 
their identifiers, more than $55$ ($50$) percent of them survive to
present day for the Phoenix(Aquarius) haloes. At given redshift,
survival fraction of the accreted subhalo is independent of the
parent halo mass, whilst the mass-loss of the subhalo is more
efficient in high mass haloes. These systematics results in
similarity and difference in the subhalo population in dark matter
haloes of different masses at present day.  \end{abstract}


\section{INTRODUCTION}
\label{sec:intro}
In the standard $\Lambda$CDM cosmology, dark matter subhaloes are
consequence of hierarchical clustering of dark matter haloes. During
the hierarchical process, the accreted dark matter halo often survive
as self-bound subhalo orbiting its host
\citep[e.g.][]{tormen98,ghigna00,springel01, delucia04, gao04, gao12,
  diemand07, springel08a}. In observations, subhaloes have been
detected with gravitation lensing \citep{vegetti12,li14} in recent
years.

Thanks to great advance in high resolution cosmological simulations,
properties of subhaloes have been extensively investigated in recent
years. Regardless of a variety of different definitions of subhalo in
cosmological simulations, numerical studies tend to agree on a
number of basic properties of the subhalo population in $\Lambda$CDM
haloes. 1) Because of efficient tidal stripping of the subhalo
population, in particularly in inner region of its host halo, the
subhalo population is a biased tracer of dark matter distribution of
its host. The distribution of subhaloes is substantially less
concentrated than that of the underlying dark matter
\citep[e.g.][]{ghigna00,  gao04,nagai05}. With ultra high
resolution of numerical simulation of the Phoenix and the Aquarius
projects, \cite{gao12} shows that the radial distribution of subhalo
is independent on the subhalo and the host halo mass.  2) The mass
function of subhaloes follow a power law relation $\dd N(>M_{sub})/\dd
M_{sub} \propto M_{sub}^{\alpha}$ \citep{boylankolchin09, gao11} with
a slope $\alpha$ varying from $1.9$ to $2$, depending on the employed
subhalo finder\citep{onions13}. The subhalo mass function is found to
be correlated with the host halo mass, with the more massive halo
tends to contain more abundant
subhaloes\citep{gao04,gao11,ishiyama13}. On average, the amplitude
of the subhalo mass of rich cluster sized haloes is about $40$ percent
higher than that of galactic haloes \citep{gao12}. The subhalo mass
function also correlates with the host halo properties, for instance
halo concentration and formation time
\citep{gao04,gao11,contini12}. Convincing and explicit explanations of
the host halo mass and properties dependence of the subhalo mass
function still lack. In light of numerical works, properties of the
subhalo population have also been extensively studied with
semi-analytic models
\citep[e.g.][]{taylor05,zentner05,giocoli08,yang12,jiangfz14}.

Most previous studies on the subject have focused on investigating the
subhalo population at redshift $z=0$. In this study, we complement those
by investigating the evolution of subhalo population. More
specifically, we will study systematics in the assembly of subhaloes
across cosmic time and in haloes of different masses. To this end, we
make use of two ultra high resolution of N-body simulations of
individual dark matter haloes from the Phoenix and the Aquarius
project. The simulated dark matter haloes in the Phoenix and Aquarius
project differ by a factor of $1000$ in the halo mass, thus provide an
ideal sample to study the assembly of subhaloes in the dark matters
halo with different masses.  It is worthwhile mentioning that the
Phoenix and the Aquarius simulations have very similar effective mass
and force resolution in terms of the simulated particle number. This
allows us to facilitate easy comparison between the two simulation
sets.

Our paper is organized as follows. In Section 2, we briefly describe
the numerical simulations used in this study. In Section 3 we present
result of the progenitor population of the Phoenix and the Aquarius
haloes before accretion. We contrast results for the evolution of the
subhalo population in the Phoenix and Aquarius simulations in Section
4. Section 5 summarize our main findings.

\section{SIMULATION}
\label{sec:simu}
Numerical simulations used in this study comprise two sets of ultra
high resolution re-simulation of individual dark matter haloes from
the Phoenix \citep{gao12} and the Aquarius Projects \citep{springel08a}
of the Virgo consortium. In terms of the numerical resolution,
the two projects are respective representation of the current
state-of-art $N$-body simulations of rich cluster and Milky way sized
dark matter haloes. For objective of numerical convergence study, both
the Phoenix and the Aquarius suits have run simulations with various
resolutions. In this study we have adopted level-2 resolution of each
simulation sets. At the level-2 resolution, each of $9$ Phoenix
clusters and $6$ Aquarius galactic haloes contains about $10^8$
particles within their virial radius $R_{200}$. Here $R_{200}$ is
defined as a radii at which the enclosed density is $200$ times of
critical density of the Universe. Hence both simulation suits indeed
have identical effective mass and force resolution. This allows us to
facilitate easy comparison of results between two simulation
sets. Note, we adopt $7$ Phoenix clusters which there is no ambiguity
in constructing their main branch merger trees.

The Phoenix cluster and the Aquarius galaxy sample were selected
for resimulation from the Millennium simulation \citep{springel05}.
The Millennium simulation assume Cosmological parameters consistent
with first year WMAP data were adopted, $\omegam =0.25, \omegab=0.045,
\omegal =0.75, h=0.73, \sigma_{8}=0.9, n=1$. These parameters deviate
from the latest CMB result, however the small offset has no
consequence for the topic addressed here. We refer readers to
\citet{gao12} and \citet{springel08a} for details of the Phoenix and
the Aquarius simulation suits.

The dark matter haloes in our simulations are identified with standard
friends-of-friends group algorithm with a linking length $0.2$ times
inter-particles separation \citep{davis85}. Based upon FOF group
catalog, we identify locally over-dense and self-bound subhaloes
with SUBFIND \citep{springel01b}. The subhalo catalog is used to
construct merger trees tracking subhaloes between snapshots
(e.g. Boly-choin et al. 2009). 

\section{The un-evolved subhalo mass function}
\label{sec:mf}
The un-evolved subhalo mass function describes the mass spectrum of
the building-up progenitors over the entire life time of a halo
assembly. It is therefore interesting to investigate whether
the un-evolved subhalo population in the cluster and the galaxy sized
halo is different. Namely whether the different subhalo abundance
between the cluster sized and galactic haloes seen today is set at the
first place. Previous studies on the subject claimed that the
un-evolved subhalo mass function follows an universal function and is
independent of halo mass (see \cite{giocoli08,lm09},
\citet{vandenbosch05}). This is surprising because the standard 
$\Lambda$CDM power spectrum is not scale free, it is not obvious that 
the un-evolved subhalo mass function should be identical in haloes 
of different scales. We re-examine this independently in this work as follows.

We first construct the halo main branch for each individual
halo. Starting from the final halo at $z=0$, we trace its most massive
progenitor in the adjacent snapshot at earlier epoch. The procedure
is repeated until the simulation lost its resolution to identify a
halo (32 dark matter particles for a FOF halo by our definition). Next
we add a halo as a progenitor candidate if it is accreted into
$R_{200}$ of its main branch at later time. The accretion time for a
subhalo is defined at the time when it has peak virial mass
 $M_{200}$ in its growth history. Correspondingly, its mass at
accretion time is defined as the mass of the progenitor halo. The
definition here is used because the stellar mass of satellite galaxies
are more tightly related to the peak $M_{200}$
  \citep{guo10, watson13,boylankolchin09}. Note there are a couple of
definitions on accretion time of a subhalo. For example, some studies
\citep{gao04, lm09} define the accretion time when an
individual halo becomes a subhalo of a FOF halo. Some studies 
\citep{giocoli08} adopt the time when an individual halo pass 
virial radius of its host halo. 

The merging history of dark haloes are generally quite complicated
\citep{kravtsov04a}. We need to consider two special cases
below. 1) 'Ejected' halo: namely a halo was only a temporal
progenitor at earlier time but passed through its host and became an
isolated haloes in its later evolution. The ejected progenitor
population have been investigated by \citep{ludlow09,wanghy09,li13}.
As these ejected progenitors have no influence on the final subhalo
population at present day, we remove them from our progenitor
catalog. 2) 'Re-accreted' halo: a 're-accreted' progenitor refer 
to a halo passing through the main branch more than once in the 
past, but either is completely disrupted or retains as a subhalo of 
the host halo lastly.  In this case, we only register it at the 
first infall. Lastly, we also add haloes merged with other 
progenitor rather than the main branch. For these progenitors, 
we apply for the same procedure above for the progenitors of the 
main branch to make sure they are members of the final halo.

Our definition of the un-evolved subhalo mass function is somewhat
different from \citet{giocoli08,giocoli10} and \citet{lm09}. In 
these works, a halo is called a progenitor if it becomes a member 
of the FOF group \citep{lm09} of the main branch haloes or pass 
through the virial radius\citep{giocoli08,giocoli10}. Whilst
we consider a halo to be a progenitor only if it is accreted into
$R_{200}$ of the host.

There are also significant differences in the detailed tracking
procedure. In \citet{giocoli08}, the authors only considered
progenitors merged with main branch but neglected the subhalo
population merged with subbranch progenitors. This of course excludes
a significant progenitor population. \citet{lm09} improved this by
considering all mergers as we did. But we differ in definition of the
progenitor masses, also in treating the 're-accreted' and the 
'ejected' halos. Since an 're-accreted' progenitor is basically 
the same object during multiple mergers with its host halo and 
contains the same central galaxy, it is more reasonable to consider 
it as a single progenitor. While in \cite{lm09}, it was repeatedly 
counted as new individual progenitors during each merging events. 
In fact, as shown below both the 're-accreted' and the 'ejected' 
halos are substantial populations of the progenitor as a whole, 
which greatly affect the un-evolved subhalo mass function.

\begin{figure}
\includegraphics[width=0.5\textwidth]{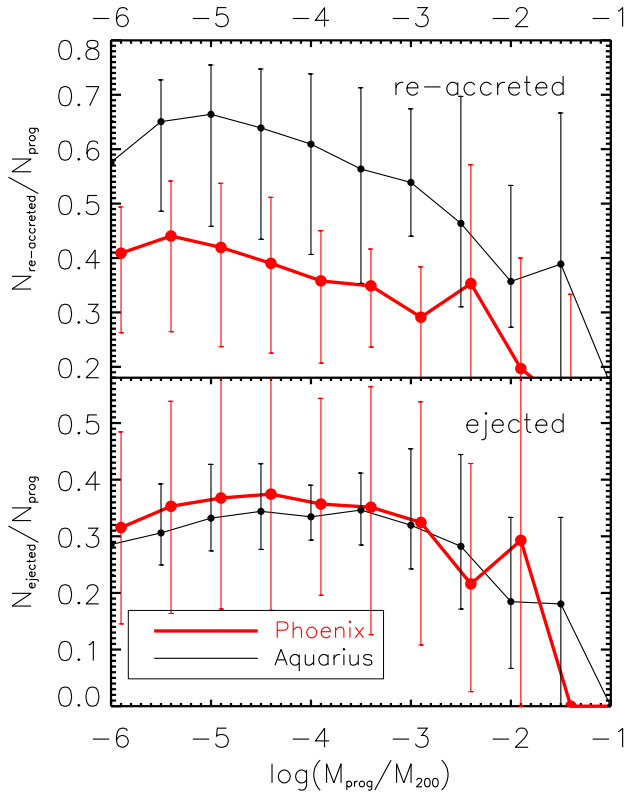}
\caption{The averaged fraction of the 're-accreted' progenitor 
(top panel) and the 'ejected' halos (bottom panel) compared to 
the total number of real progenitors as a function of progenitor 
halo mass. The results of $7$ Phoenix halos are shown as thick 
red lines. The results if $6$ Aquarius halos are shown as thin
 black lines. The error bars on selected points show full
scatters of our samples.}
\label{fig:ejected}
\end{figure}

In the top panel of Figure~\ref{fig:ejected}, we plot the averaged
fraction of the 're-accreted' progenitor of the $7$ Phoenix and the
$6$ Aquarius haloes as a function of the progenitor mass (normalized
to the host halo mass at $z=0$) by solid lines.  Only for the massive
progenitors with mass greater than $1/100$ of their parent, the
fraction of the 're-accreted' progenitors is small for both the
Phoenix and the Aquarius haloes. However, for progenitors less massive
than the ratio, the 're-accreted' progenitor is a significant 
population of the progenitor as a whole. About $40$ percent of
progenitors are  're-accreted' progenitors in Phoenix clusters. Up
to $60$ percent progenitors of Aquarius halos are 're-accreted'
progenitors. The 're-accreted' fraction decreases with increasing
progenitor mass.  These 're-accreted' progenitors have been
repeatedly counted in un-evolved subhalo mass function of 
\citet{giocoli08} and \citet{lm09}.

The bottom panel of figure~\ref{fig:ejected} shows 
the averaged ratio between the number of the 'ejected' halos and
real progenitors as a function of the progenitor mass. The 'ejected'
fraction is about $30\%-35\%$ and seems independent on progenitor and
host halo mass. 

\begin{figure}
\includegraphics[width=0.5\textwidth]{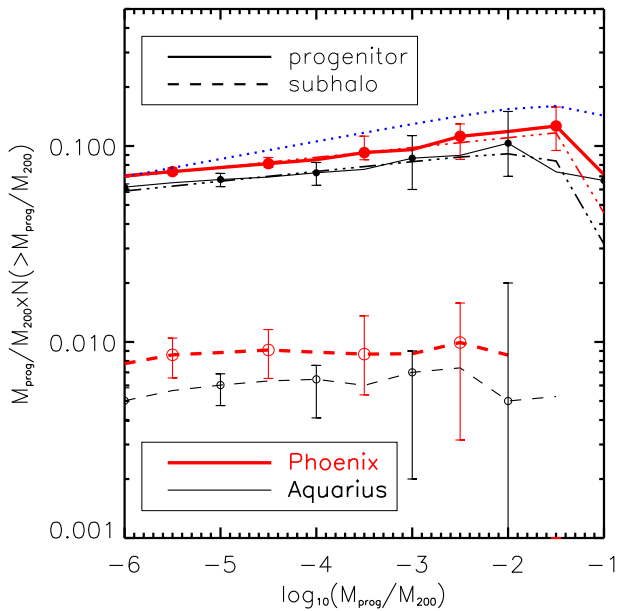}
\caption{Cumulative un-evolved (solid lines) and present subhalo mass
  function(dashed lines) of cluster sized Phoenix (red, thick) and galactic
  Aquarius (black, thin) haloes. The $y$ axis has been multiplied by the
  normalized mass in order to expand the dynamic range. The lines show
  the median subhalo mass function for samples of $7$ haloes in the
  Phoenix and the $6$ haloes in the Aquarius simulation suits. The
  error bars show whole scatter about the median. 
  The dash-dotted lines show our fits to our simulations.
  The un-evolved subhalo mass function of \protect\cite{lm09} is shown 
  as blue dotted lines.}
\label{fig:mf}
\end{figure}

In Figure~\ref{fig:mf} we plot our own un-evolved subhalo mass function
for the Phoenix and the Aquarius simulation suits. In the plot, the
median value of the cumulative un-evolved subhalo mass function of the $7$
Phoenix and the $6$ Aquarius haloes are shown as red and black solid
lines, respectively. The cumulative mass functions are multiplied by
$M_{prog}/M_{halo}$ in order to remove the dominant mass dependence and
make the differences between curves more apparent. Clearly the
un-evolved subhalo mass function depends on the halo mass, with the
cluster sized Phoenix having $20\%$ more progenitors than that of
Aquarius galactic haloes. 
The un-evolved  mass functions are well fitted by:
\begin{equation}
  f(N>\mu \equiv \frac{M_{prog}}{M_{200}})=(\frac{\mu}{a})^b \rm{exp}[-(\frac{\mu}{c})^d]
\end{equation}
  For Phoenix haloes, the parameters are $a=0.125, b=-0.95, c=0.1, d=12$.
  For Aquarius haloes, the parameters are $a=0.105, b=-0.95, c=0.09,d=1.72$.

Our result hence is inconsistent with
\citet{giocoli08} and \citet{lm09}, who claimed that an universal form
for the un-evolved subhalo mass function independent of the host halo
mass. The reason for the discrepancy  has been discussed above when we
elucidate differences in the definitions of the un-evolved subhalo 
population among ours. It is worthwhile mentioning that when we use the 
same definition of the un-evolved subhalo population as 
\citet{giocoli08,lm09}, our own results agree quite well with 
\citet{giocoli08} and \citet{lm09}. Hence the discrepancy shown here is 
entirely because of the different definitions of the un-evolved subhalo 
population. For comparison, we over-plot the fit of all order un-evolved 
subhalo mass function of \citet{lm09} as a blue dotted curve The 
amplitude is about $20\%$ to $50\%$ lager than ours. This is expected 
for three reasons. Firstly, \citet{lm09} based their work on FOF haloes, 
whereas ours is based upon spherical over-densities. Secondly 
're-accreted' progenitors were counted multiple times by \citet{lm09}, 
and 'ejected' haloes were included by \citet{lm09}, these increase the 
amplitude of the un-evolved  subhalo mass function. 
Third, we use peak $M_{200}$ as infall mass, which is larger than that 
defined in \citet{lm09}.

For easy reference, we over-plot the median of the subhalo mass
function of the Phoenix and Aquarius haloes as black and red dashed
lines in the same figure, respectively.  The difference in the
amplitude of the subhalo mass function between the Phoenix and the
Aquarius is about $40$ percent, larger than the difference seen in the
un-evolved subhalo mass function. This suggests that subsequent
evolution of the subhalo population should also be important to
account for the parent halo mass dependent subhalo mass function
abundance as will be analyzed below.

\section{The assembly history of subhalo population}
\label{sec:massdependence}

Above we studied the progenitor population before accretion. In this
section, we will investigate the evolution of the progenitors after
they were accreted into their parent haloes.

\subsection{Accretion time distribution of progenitors}
\label{subsec:at}
\begin{figure}
\includegraphics[width=0.5\textwidth]{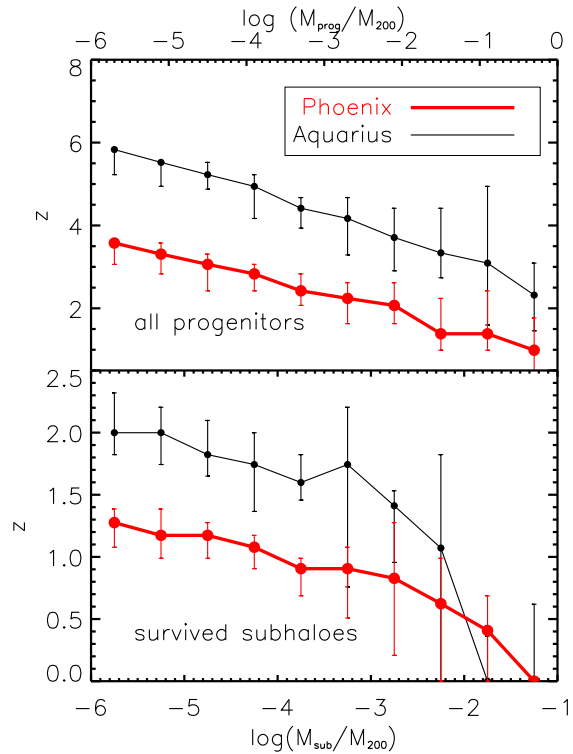}	
\caption{Accretion time distribution. The top panel show the median 
  accretion time distribution of all progenitors of the Phoenix and the 
  Aquarius haloes as a function of the progenitor mass (upper axis). The
  bottom panel show the median accretion time  of the survived subhaloes
  as a function of the subhalo mass. The median values of the Phoenix
  and the Aquarius haloes are plotted. Different colours are used to
  distinguish the different simulation sets as indicated in the
  legend. The error bars show the full scatter about the median.}
\label{fig:mprog_a}
\end{figure}

In top panel of Fig:~\ref{fig:mprog_a}, we show the accretion time
distribution of all progenitors as a function of the progenitor
mass (upper axis) for the Phoenix and Aquarius haloes. Hereafter we
will use the normalized mass for the progenitor and the subhalo rather
than their actual masses in order to take out the host halo mass
dependence. The median accretion time of all progenitors in the entire
assembly history of $7$ phoenix and $6$ Aquarius haloes is shown as
the red and black solid lines, respectively. Error bars display full
scatter of the accretion time distribution for the Phoenix and Aquarius
haloes. Here we define the accretion time as the redshift when a
progenitor reaches its peak $M_{200}$. Clearly, there is a strong
relation between the accretion time and the progenitor mass for both
the Phoenix and the Aquarius haloes, with less massive progenitors
accreted earlier than their more massive counterparts. Most Progenitors of
the Aquarius haloes are typically accreted before redshift $5$, while
it is about $z\sim 3$ for the Phoenix haloes. 

The offset in the accretion time distribution between the Phoenix and the Aquarius
haloes may reflect the fact that cluster haloes are assembled later.

The bottom panel of Fig:~\ref{fig:mprog_a} show the accretion time
distribution of the present day survived subhaloes as a function of
the subhalo mass (lower axis). It follows the same trend as that of
the progenitor population, with less massive subhaloes accreted
earlier. As it can been clearly seen that most subhaloes of galactic
haloes were accreted before redshift $1.7$, whilst it is before redshift
$0.9$ for the Phoenix subhaloes. The result is consistent with the study
of \citet{boylankolchin09} who used the same definition of the
accretion time to ours, while we extend the result to a lower subhalo
mass. However this result is inconsistent with an earlier work of
\citet{gao04} who found that most subhaloes are accreted later
than $z=0.5$. The discrepancy mainly lies in the definition of accretion
time. In \citet{gao04}, the accretion time of a subhalo is defined as
the time it was lastly associated to an individual FOF halo. As we
showed in the previous section that a quite large population of the
're-accreted' progenitor repeatedly passed through and was later
re-accreted. In this work, the accretion time is defined at the time 
when a progenitor  achieves its peak $M_{200}$, which is substantially earlier
\citep{behroozi14} than that of \citet{gao04}.

\subsection{The fate of accreted progenitors}
\label{subsec:rf}
After accretion, some progenitors will survive as subhaloes at present
day, some will be be completely distroyed by tidal field. Below we
investigate the survival ability of progenitors. What is the
fraction of them survive to present day? For the progenitor population
accreted at a fixed redshift, what is the surviving fraction today?
How much mass is retained in the survived subhaloes?  It is also
interesting to study whether the above physical quantities depends on
the host halo mass.

\begin{figure}
\includegraphics[width=0.5\textwidth]{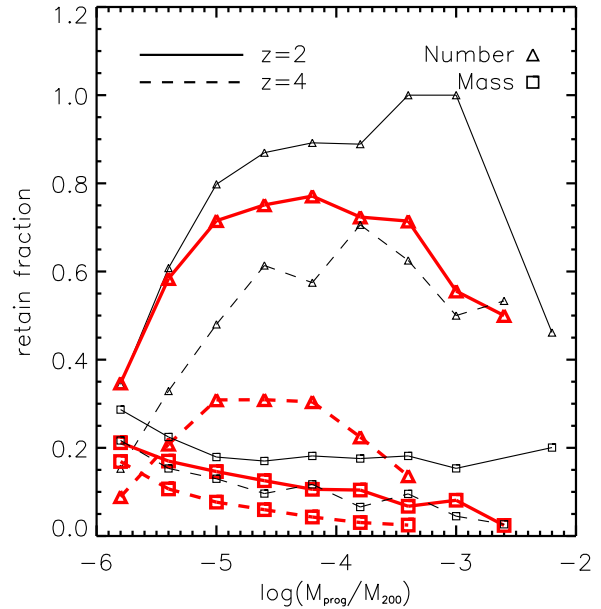}
\caption{The survival number and mass fraction of progenitors accreted
  at redshift 4 and 2 as a function of the progenitors mass. The solid
  lines show result for redshift 2, and the dashed lines show result
  for 4. The number and mass fraction are distinguished with triangles
  and squares, respectively. Thick red curves show results for
  the Phoenix. Thin black curves are for the Aquarius.}
\label{fig:retain}
\end{figure}

We firstly consider below the survival number and mass fraction of
progenitors who were accreted at two fixed redshifts, $z=2$ and
$4$. Results are shown in Figure~\ref{fig:retain}. The triangles
connected by lines represent the survival number fraction of
progenitors, while the squares connected by lines are for the retained
mass fraction. The result for the Phoenix and the Aquarius haloes are
distinguished with different colours. For massive progenitors with
normalized mass $log(M_{prog}/M_{200,z=0})>10^{-5}$ at $z=2$,  more
than $80$ percent of the entire progenitor population survives to the
present day in  Aquarius haloes. The fraction is $70$ percent for 
Phoenix haloes. The survival number fraction is independent on the 
progenitor mass.
 The decline in the survival number fraction at the low mass end may
be because of limited numerical resolution of our
simulations. 
The retained mass fraction is quite different between two simulations. 
For progenitors of the Aquarius haloes, about $20$ 
percent of its original mass is retained, a factor of $2$ larger 
than those of Phoenix haloes. This suggests that tidal stripping 
process is more efficient in cluster than in galactic environments.
 Results for the progenitors accreted at $z=4$ are qualitatively
similar, albeit both the survival number and mass fraction is slightly
lower because of earlier infall. Still about $60$ percent of the
accreted progenitors survive as entities in the tidal disruption
process in Aquarius haloes. In Phoenix haloes the survival number 
fraction is much lower, at most $30$ percent of progenitors survive 
through the tidal disruption. About $5-15$  percent of their original 
mass is retained. The survival fraction and retain mass fraction might be 
slightly underestimated at massive end, since SUBFIND we used has trouble 
to identify subhalos near halo center\citep{muldrew11}.

\begin{figure}
\includegraphics[width=0.5\textwidth]{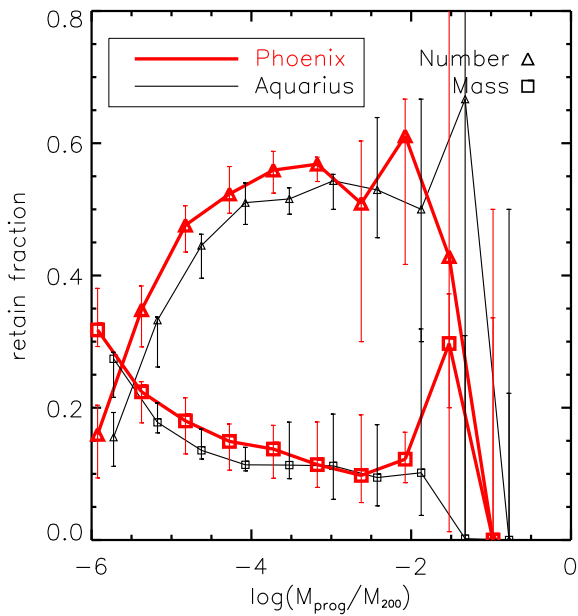}
\caption{The survival number and mass fraction of all progenitors as a
  function of the progenitor mass. The triangles are the survived
  number fraction, the squares are the survived mass fraction. Thick red
curves show results for the Phoenix. Thin black curves are for
the Aquarius. The error bars show the full scatter of median.}
\label{fig:retain_all}
\end{figure}

In Figure~\ref{fig:retain_all} we plot the survival number and mass
fraction of all progenitors as a function of the progenitor mass
during the entire assembly history of the Phoenix and the Aquarius
haloes, where the median values of the Phoenix and the Aquarius haloes
are shown. Different colours are used to distinguish different 
simulation sets, and different symbols are used to distinguish the
number and mass fraction. As can be seen clearly that the survival number
fraction of progenitors depend strongly on the progenitor
mass for progenitors more massive than $1/100$ of their host halo
mass,  with more massive progenitors more easily destroyed. This is
expected because dynamical friction effect is stronger for massive
subhaloes, which assists the tidal stripping. Once progenitors  mass
are less than $1/100$ of their hosts, the survival number fraction
becomes largely independent of the progenitor mass. During the entire
assembly of Phoenix haloes, roughly $55$ percent progenitors survive
as subhaloes at present day, the survival fraction is only slightly
lower for Aquarius haloes, which is about $50$ percent. Presumably
this results from earlier accretion of progenitor in galactic Aquarius
haloes as we discussed above. Towards the low progenitor mass end,
there is a drop in the survival number fraction, this is very likely
caused by the numerical resolution of our simulations. 

Square symbols connected with solid lines show the survival mass
fraction of Phoenix and Aquarius haloes. In spite of the noticeable
difference in the survival number fraction between two simulation
sets, the survival mass fraction of both simulations are quite
similar. While most progenitors survive as entities at present day,
about $90$ percent of their original mass is striped.  Although the
accretion time of progenitors is more recent for Phoenix haloes, their
survival mass fraction is very similar to that of the Aquarius,
reflecting the fact that the tidal disruption process is stronger in
more massive systems, consistent with what shows in the
Figure~\ref{fig:retain}.

\subsection{Radial dependence of the retained mass of progenitors}
\label{subsec:ss}

A subhalo orbiting within its parent for a long time will have
suffered significantly  from the effects of dynamical friction and
tidal stripping, so its orbit will have decayed by a larger factor
than that of a recently accreted subhalo of similar current mass. 
This effect is expected to result in  a correlation between the radial
position of a subhalo and its accretion time. \citet{gao04} showed
that there is indeed a tight relation between the retained mass
fraction of progenitors and their radial position (see also
\cite{kravtsov04b}). The relation can be tested in future
Galaxy-Galaxy lensing  observations \citep{li13,li14} . With $1000$
times better resolution simulations, we revise the relation as
follows.

\begin{figure}
\includegraphics[width=0.5\textwidth]{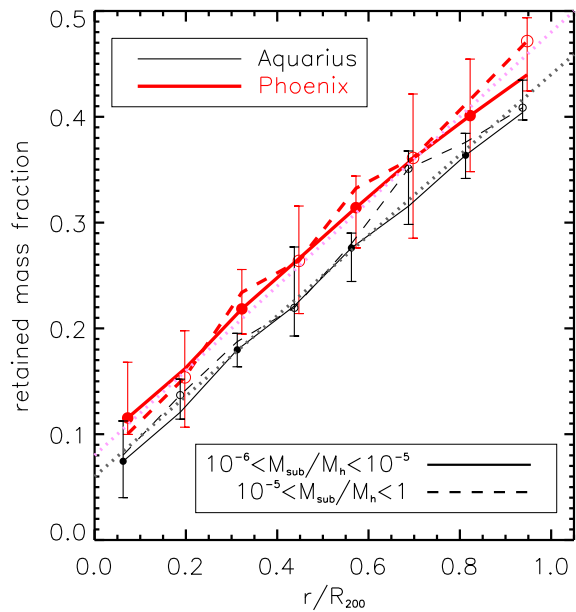}
\caption{The ratio between the present subhalo mass and its original
  mass at accretion as a function of centric distance.The medial
  value of the Phoenix and Aquarius are shown. The solid lines are the
  results of subhaloes with present mass $M_{sub}/M_{h}<10^{-5}$. The
  dashed lines are the results of subhaloes with present mass
  $M_{sub}/M_{h}>10^{-5}$. Thick red curves show results for the Phoenix
  and thin black curves are for the Aquarius. The blue dotted lines
  show a linear fits to the radial dependence.}
\label{fig:radius}
\end{figure}

In Figure~\ref{fig:radius} we plot the median values of the retained
mass fraction of subhaloes against $r/R_{200}$ for subhaloes of the
Phoenix and the Aquarius haloes. In order to investigate whether the
relation depends on the subhalo mass, we divide our subhalo population
into two sub-samples, $10^{-6}< M_{sub}/M_{h}<10^{-5}$ and $10^{-5}<
M_{sub}/M_{h}$, respectively. The error bars represent full scatters
of the Phoenix and the Aquarius suits. The strong radial dependence of
retained mass fraction seems largely independent of the subhalo mass. 
It weakly depends on the host halo mass, $4$ percent difference in mass,
 but with large scatters.  We provide a linear fit of a form:
\begin{equation}
f(r/R_{200})=a\times r/R_{200}+b
\end{equation}
For Phoenix haloes, $a=0.4, b=0.08$. For Aquarius haloes, $a=0.38, b=0.06$.
The fits are shown as pink (Phoenix) or gray (Aquarius) dashed lines.

\section{Conclusion}
We take advantage of two sets of ultra high resolution $N$-body
simulations from the Phoenix and the Aquarius Project to explore
systematics in the assembly history of subhaloes in dark matter haloes
of different masses. The Phoenix and the Aquarius simulations have
the same effective mass and force resolution. This allows us to make
fair comparison of the evolution of subhaloes and their progenitors
between cluster and galaxy sized dark matter haloes. 

We have adopt a more detailed tracking procedure to follow the
evolution of progenitors. By exercising this, we find that a quite large
population of progenitors of a final host halo passed through and
were later re-accreted to the main progenitor more than once. The
averaged fraction of the 're-accreted' progenitors vary with the halo
mass, which is about $60$ percent for the Aquarius haloes and $40$
percent for the Phoenix haloes.There is also a substantial 'ejected' 
halo population accreted to its host in the past but being in 
isolation at present day. We find that the abundance of progenitors 
which build up present day subhalo population systematically depends 
on the host halo  mass, in contrast to previous studies. The 
amplitude of the un-evolved subhalo mass function depends 
systematically on the host halo mass, the cluster sized haloes
on average have at least $20$ percent more progenitors than that of
galactic counterparts. 

The accretion time of progenitors depend on the host halo mass as
well as the progenitor mass. Typically most progenitors of the
galactic haloes were accreted before a redshift $z=5$, while the
accretion time for progenitors of clusters is about $z=3$. Less
massive progenitors are accreted earlier than more massive ones. At
the fixed progenitor mass, the survival number fraction of progenitors
does not depend on the host halo mass, while the retained mass
fraction correlates with the host halo mass. Tidal striping is more
efficient in cluster environment than in that of galaxy. For the
progenitor population as a whole, $55$ percent of them are able to
survive as entities at present day for galactic haloes, the survival
number fraction for clusters is only sightly lower, which is $50$
percent. Nevertheless, the survived subhaloes roughly retain $10$
percent of their original mass, independent on the host halo mass. The
evolution of subhaloes leads to a radial dependence of the retained
mass fraction. The relation seems largely independent of the host halo
mass as well as the subhalo mass, we provide a simple fit to it.

These systematics between the assembly of subhaloes in cluster sized
and galactic haloes should account for similarities and differences in
the host mass dependence of subhalo population seen in N-body
Cosmological simulations.

\section*{Acknowledgements}
Phoenix and Aquarius are projects of the Virgo Consortium. Most
simulations were carried out on the Lenova Deepcomp7000 supercomputer
of the super Computing Center of Chinese Academy of Sciences, Beijing,
China, and on Cosmology machine at the Institute for Computational
Cosmology (ICC) at Durham. The Cosmology machine is part of the
{\small DiRAC} facility jointly founded by {\small STFC}, the large
facilities capital fund of {\small BIS}, and Durham University. LG
acknowledges support from {\small NSFC} grants (Nos. 11133003 and
11425312), {\small MPG} partner Group family, and an {\small STFC}
Advanced Fellowship,  as well as the hospitality of the Institute for
Computational Cosmology at Durham University. 
 
\bibliographystyle{mn2e}
\setlength{\bibhang}{2.0em}
\setlength\labelwidth{0.0em}
\bibliography{smf}
\label{lastpage}
\end{document}

%% file: macros_xie.tex
\newcommand{\dd}{{\rm d}}

\newcommand{\omegam}{\Omega_{\rm m}}
\newcommand{\omegal}{\Omega_{\rm \Lambda}}
\newcommand{\omegab}{\Omega_{\rm b}}

\def\gsim { \lower .75ex \hbox{$\sim$} \llap{\raise .27ex \hbox{$>$}}}
\def\lsim { \lower .75ex \hbox{$\sim$} \llap{\raise .27ex \hbox{$<$}}}

%% file: smf.bbl
\begin{thebibliography}{36}
\expandafter\ifx\csname natexlab\endcsname\relax\def\natexlab#1{#1}\fi

\bibitem[{{Behroozi} {et~al}\mbox{.}(2014){Behroozi}, {Wechsler}, {Lu}, {Hahn},
  {Busha}, {Klypin}, \& {Primack}}]{behroozi14}
{Behroozi} P.~S., {Wechsler} R.~H., {Lu} Y., {Hahn} O., {Busha} M.~T., {Klypin}
  A., {Primack} J.~R., 2014, \apj, 787, 156

\bibitem[{{Boylan-Kolchin} {et~al}\mbox{.}(2009){Boylan-Kolchin}, {Springel},
  {White}, {Jenkins}, \& {Lemson}}]{boylankolchin09}
{Boylan-Kolchin} M., {Springel} V., {White} S.~D.~M., {Jenkins} A., {Lemson}
  G., 2009, \mnras, 398, 1150

\bibitem[{{Contini}, {De Lucia} \& {Borgani}(2012){Contini}, {De Lucia}, \&
  {Borgani}}]{contini12}
{Contini} E., {De Lucia} G., {Borgani} S., 2012, \mnras, 420, 2978

\bibitem[{{Davis} {et~al}\mbox{.}(1985){Davis}, {Efstathiou}, {Frenk}, \&
  {White}}]{davis85}
{Davis} M., {Efstathiou} G., {Frenk} C.~S., {White} S.~D.~M., 1985, \apj, 292,
  371

\bibitem[{{De Lucia} {et~al}\mbox{.}(2004){De Lucia}, {Kauffmann}, {Springel},
  {White}, {Lanzoni}, {Stoehr}, {Tormen}, \& {Yoshida}}]{delucia04}
{De Lucia} G., {Kauffmann} G., {Springel} V., {White} S.~D.~M., {Lanzoni} B.,
  {Stoehr} F., {Tormen} G., {Yoshida} N., 2004, \mnras, 348, 333

\bibitem[{{Diemand}, {Kuhlen} \& {Madau}(2007){Diemand}, {Kuhlen}, \&
  {Madau}}]{diemand07}
{Diemand} J., {Kuhlen} M., {Madau} P., 2007, \apj, 667, 859

\bibitem[{{Gao} {et~al}\mbox{.}(2011){Gao}, {Frenk}, {Boylan-Kolchin},
  {Jenkins}, {Springel}, \& {White}}]{gao11}
{Gao} L., {Frenk} C.~S., {Boylan-Kolchin} M., {Jenkins} A., {Springel} V.,
  {White} S.~D.~M., 2011, \mnras, 410, 2309

\bibitem[{{Gao} {et~al}\mbox{.}(2012){Gao}, {Navarro}, {Frenk}, {Jenkins},
  {Springel}, \& {White}}]{gao12}
{Gao} L., {Navarro} J.~F., {Frenk} C.~S., {Jenkins} A., {Springel} V., {White}
  S.~D.~M., 2012, \mnras, 425, 2169

\bibitem[{{Gao} {et~al}\mbox{.}(2004){Gao}, {White}, {Jenkins}, {Stoehr}, \&
  {Springel}}]{gao04}
{Gao} L., {White} S.~D.~M., {Jenkins} A., {Stoehr} F., {Springel} V., 2004,
  \mnras, 355, 819

\bibitem[{{Ghigna} {et~al}\mbox{.}(2000){Ghigna}, {Moore}, {Governato}, {Lake},
  {Quinn}, \& {Stadel}}]{ghigna00}
{Ghigna} S., {Moore} B., {Governato} F., {Lake} G., {Quinn} T., {Stadel} J.,
  2000, \apj, 544, 616

\bibitem[{{Giocoli} {et~al}\mbox{.}(2010){Giocoli}, {Tormen}, {Sheth}, \& {van
  den Bosch}}]{giocoli10}
{Giocoli} C., {Tormen} G., {Sheth} R.~K., {van den Bosch} F.~C., 2010, \mnras,
  404, 502

\bibitem[{{Giocoli}, {Tormen} \& {van den Bosch}(2008){Giocoli}, {Tormen}, \&
  {van den Bosch}}]{giocoli08}
{Giocoli} C., {Tormen} G., {van den Bosch} F.~C., 2008, \mnras, 386, 2135

\bibitem[{{Guo} {et~al}\mbox{.}(2010){Guo}, {White}, {Li}, \&
  {Boylan-Kolchin}}]{guo10}
{Guo} Q., {White} S., {Li} C., {Boylan-Kolchin} M., 2010, \mnras, 404, 1111

\bibitem[{{Ishiyama} {et~al}\mbox{.}(2013){Ishiyama}, {Rieder}, {Makino},
  {Portegies Zwart}, {Groen}, {Nitadori}, {de Laat}, {McMillan}, {Hiraki}, \&
  {Harfst}}]{ishiyama13}
{Ishiyama} T. {et~al.}, 2013, \apj, 767, 146

\bibitem[{{Jiang} \& {van den Bosch}(2014)}]{jiangfz14}
{Jiang} F., {van den Bosch} F.~C., 2014, ArXiv: 1403.6827

\bibitem[{{Kravtsov} {et~al}\mbox{.}(2004){Kravtsov}, {Berlind}, {Wechsler},
  {Klypin}, {Gottl{\"o}ber}, {Allgood}, \& {Primack}}]{kravtsov04a}
{Kravtsov} A.~V., {Berlind} A.~A., {Wechsler} R.~H., {Klypin} A.~A.,
  {Gottl{\"o}ber} S., {Allgood} B., {Primack} J.~R., 2004, \apj, 609, 35

\bibitem[{{Kravtsov}, {Gnedin} \& {Klypin}(2004){Kravtsov}, {Gnedin}, \&
  {Klypin}}]{kravtsov04b}
{Kravtsov} A.~V., {Gnedin} O.~Y., {Klypin} A.~A., 2004, \apj, 609, 482

\bibitem[{{Li} {et~al}\mbox{.}(2013){Li}, {Mo}, {Fan}, {Yang}, \&
  {Bosch}}]{li13}
{Li} R., {Mo} H.~J., {Fan} Z., {Yang} X., {Bosch} F.~C.~v.~d., 2013, \mnras,
  430, 3359

\bibitem[{{Li} {et~al}\mbox{.}(2014){Li}, {Shan}, {Mo}, {Kneib}, {Yang}, {Luo},
  {van den Bosch}, {Erben}, {Moraes}, \& {Makler}}]{li14}
{Li} R. {et~al.}, 2014, \mnras, 438, 2864

\bibitem[{{Li} \& {Mo}(2009)}]{lm09}
{Li} Y., {Mo} H., 2009, ArXiv :0908.0301

\bibitem[{{Ludlow} {et~al}\mbox{.}(2009){Ludlow}, {Navarro}, {Springel},
  {Jenkins}, {Frenk}, \& {Helmi}}]{ludlow09}
{Ludlow} A.~D., {Navarro} J.~F., {Springel} V., {Jenkins} A., {Frenk} C.~S.,
  {Helmi} A., 2009, \apj, 692, 931

\bibitem[{{Muldrew}, {Pearce} \& {Power}(2011){Muldrew}, {Pearce}, \&
  {Power}}]{muldrew11}
{Muldrew} S.~I., {Pearce} F.~R., {Power} C., 2011, \mnras, 410, 2617

\bibitem[{{Nagai} \& {Kravtsov}(2005)}]{nagai05}
{Nagai} D., {Kravtsov} A.~V., 2005, \apj, 618, 557

\bibitem[{{Onions} {et~al}\mbox{.}(2013){Onions}, {Ascasibar}, {Behroozi},
  {Casado}, {Elahi}, {Han}, {Knebe}, {Lux}, {Merch{\'a}n}, {Muldrew},
  {Neyrinck}, {Old}, {Pearce}, {Potter}, {Ruiz}, {Sgr{\'o}}, {Tweed}, \&
  {Yue}}]{onions13}
{Onions} J. {et~al.}, 2013, \mnras, 429, 2739

\bibitem[{{Springel} {et~al}\mbox{.}(2008){Springel}, {Wang}, {Vogelsberger},
  {Ludlow}, {Jenkins}, {Helmi}, {Navarro}, {Frenk}, \& {White}}]{springel08a}
{Springel} V. {et~al.}, 2008, \mnras, 391, 1685

\bibitem[{{Springel} {et~al}\mbox{.}(2005){Springel}, {White}, {Jenkins},
  {Frenk}, {Yoshida}, {Gao}, {Navarro}, {Thacker}, {Croton}, {Helly},
  {Peacock}, {Cole}, {Thomas}, {Couchman}, {Evrard}, {Colberg}, \&
  {Pearce}}]{springel05}
{Springel} V. {et~al.}, 2005, \nat, 435, 629

\bibitem[{{Springel} {et~al}\mbox{.}(2001){Springel}, {White}, {Tormen}, \&
  {Kauffmann}}]{springel01}
{Springel} V., {White} S.~D.~M., {Tormen} G., {Kauffmann} G., 2001, \mnras,
  328, 726

\bibitem[{{Springel}, {Yoshida} \& {White}(2001){Springel}, {Yoshida}, \&
  {White}}]{springel01b}
{Springel} V., {Yoshida} N., {White} S.~D.~M., 2001, \na, 6, 79

\bibitem[{{Taylor} \& {Babul}(2005)}]{taylor05}
{Taylor} J.~E., {Babul} A., 2005, \mnras, 364, 515

\bibitem[{{Tormen}, {Diaferio} \& {Syer}(1998){Tormen}, {Diaferio}, \&
  {Syer}}]{tormen98}
{Tormen} G., {Diaferio} A., {Syer} D., 1998, \mnras, 299, 728

\bibitem[{{van den Bosch}, {Tormen} \& {Giocoli}(2005){van den Bosch},
  {Tormen}, \& {Giocoli}}]{vandenbosch05}
{van den Bosch} F.~C., {Tormen} G., {Giocoli} C., 2005, \mnras, 359, 1029

\bibitem[{{Vegetti} {et~al}\mbox{.}(2012){Vegetti}, {Lagattuta}, {McKean},
  {Auger}, {Fassnacht}, \& {Koopmans}}]{vegetti12}
{Vegetti} S., {Lagattuta} D.~J., {McKean} J.~P., {Auger} M.~W., {Fassnacht}
  C.~D., {Koopmans} L.~V.~E., 2012, \nat, 481, 341

\bibitem[{{Wang}, {Mo} \& {Jing}(2009){Wang}, {Mo}, \& {Jing}}]{wanghy09}
{Wang} H., {Mo} H.~J., {Jing} Y.~P., 2009, \mnras, 396, 2249

\bibitem[{{Watson} \& {Conroy}(2013)}]{watson13}
{Watson} D.~F., {Conroy} C., 2013, \apj, 772, 139

\bibitem[{{Yang} {et~al}\mbox{.}(2012){Yang}, {Mo}, {van den Bosch}, {Zhang},
  \& {Han}}]{yang12}
{Yang} X., {Mo} H.~J., {van den Bosch} F.~C., {Zhang} Y., {Han} J., 2012, \apj,
  752, 41

\bibitem[{{Zentner} {et~al}\mbox{.}(2005){Zentner}, {Berlind}, {Bullock},
  {Kravtsov}, \& {Wechsler}}]{zentner05}
{Zentner} A.~R., {Berlind} A.~A., {Bullock} J.~S., {Kravtsov} A.~V., {Wechsler}
  R.~H., 2005, \apj, 624, 505

\end{thebibliography}
